\documentclass[12pt,prd,showpacs,tightenlines,nofootinbib]{revtex4}
\usepackage{bm}
\usepackage{graphicx}
\usepackage{rotating}
\usepackage{epsfig}
\begin{document}

\title{Strange baryon spectroscopy in the
  relativistic quark model } 

\author{R. N. Faustov}
\author{V. O. Galkin}
\affiliation{Dorodnicyn Computing Center, Federal Research Center
  ``Computer Science and Control'', Russian Academy of Sciences,
  Vavilov Str. 40, 119333 Moscow, Russia}

\begin{abstract}
Mass spectra of strange baryons are calculated in the framework of
the relativistic quark model based on the quasipotential
approach. Baryons are treated as the relativistic quark-diquark bound
systems. It is assumed that two quarks with equal constituent
masses form a diquark. The diquark excitations and its internal structure
are consistently taken into account. Calculations are performed up to
rather high orbital and radial excitations of strange baryons. On this
basis the Regge trajectories are constructed. The obtained
results are compared with available experimental data and previous predictions. It is found that all masses of the 4- and 3-star, as
well as most of the 2- and 1-star states of strange baryons with
established quantum numbers are well reproduced. The developed
relativistic quark-diquark model predicts less excited states than 
three-quark models of strange baryons.      
\end{abstract}


\maketitle

\section{Introduction}
\label{sec:intr}

At present the extensive evidence (including lattice calculations) of the existence of diquark
correlations in hadrons was collected.~\footnote{A vast literature on
  this subject is available. Therefore we mostly refer to the recent
  reviews where the references to earlier review and original papers can be found.} It continues constantly growing
with the accumulation of new data on various properties of light and
heavy hadrons \cite{pdg}. Thus recently several charged charmonium- and
bottomonium-like states were discovered \cite{pdg,qwg}. They should be inevitably
multiquark, at least four quark --- tetraquark, states. One of the
most successful pictures of such tetraquark states is the
diquark-antidiquark model \cite{maianih,htetr}. In the light meson sector
it has been argued for a long time that mesons forming the inverted
lightest scalar  nonet can be well described as 
tetraquarks  \cite{jaffe}  treated as
diquark-antidiquark bound states \cite{maianil,ltetr}. In the baryon
sector it is well known that the number of observed excited states
both in the
light and heavy sectors is considerably lower than the number of
excited states predicted in the three-quark picture \cite{kr,cr,aamnsv,fk}. The introduction
of diquarks significantly reduces this number of baryon
states since in such a picture some of degrees of freedom are frozen
and thus the number of possible excitations is substantially smaller.  

In our previous papers  we developed the
relativistic quark-diquark model of doubly heavy \cite{dhbar} and  heavy
baryons  \cite{hbar,hbarRegge}. We assumed that two heavy quarks in a doubly heavy baryon and
two light quarks in a heavy baryon form a diquark. The relativistic
quasipotential equation with the QCD-motivated quark-quark interaction
was solved for obtaining diquark characteristics, such as the diquark
masses and form factors. The calculation of the diquark form factors is
necessary for taking into account the diquark internal structure. For doubly
heavy diquarks \cite{dhbar} we considered both ground and excited
states, while for light diquarks \cite{hbar} we limited ourselves by only ground state scalar and axial vector
diquarks. Then the baryon masses were calculated by solving the
relativistic quark-diquark equation. It was found that the heavy
baryon spectra are well described in the proposed approach
\cite{hbar,hbarRegge}. The calculated baryon wave functions were used
for the description of  weak decays of the doubly heavy and heavy
baryons in Refs.~\cite{dhbdecay,hbardecay}.       

Here we extend our relativistic quark-diquark model for the calculation
of the mass spectra of strange baryons. These baryons are considered
as the bound systems of a quark and diquark, where we assume that a
diquark is composed from quarks of the same constituent mass. Thus $\Lambda$
and $\Sigma$ baryons contain the strange $s$ quark and the light $qq$
($q=u,d$) diquark, while $\Xi$ and $\Omega$ baryons contain the light $q$ or
strange $s$ quark and the strange $ss$ diquark. Our analysis of the strange
baryon spectroscopy shows that it is necessary to consider both ground and
excited states of these diquarks. As the result the number of obtained
baryon states is increased, however it is still significantly less
than in the three-quark approaches. The differences become evident for
the higher quark excitations in a baryon.  Our goal is to calculate the
strange baryon spectra up to rather high orbital and radial
excitations. On this basis the Regge trajectories for these baryons
can be constructed and their linearity can be tested. Moreover, the
comparison of the Regge trajectory slopes for strange and charmed
baryons as well as light mesons can be made.

The paper is organized as follows. In Sec.~\ref{sec:rqdm} we briefly
describe our relativistic quark-diquark model of baryons. Expressions
for the quasipotentials of the quark-quark interaction in a diquark and
the quark-diquark interaction in a baryon are given which include both the
spin-independent and spin-dependent relativistic contributions. Masses and form
factor parameters of ground and excited states of diquarks are
calculated. In Sec.~\ref{sec:sbmass} the mass spectra of strange baryons are
considered. The obtained  results are confronted with available
experimental data and predictions of other approaches. We calculate
the strange baryon masses up to rather high orbital and radial
excitations in the quark-diquark bound system. This allows us to
construct  their Regge trajectories which are presented in Sec.~\ref{sec:rtsb}. The
corresponding slopes and intercepts are determined. Finally, we give
our conclusions in Sec.~\ref{sec:concl}.

\section{Relativistic quark-diquark model}
\label{sec:rqdm}

For the calculations of the strange baryon spectra we employ the quasipotential approach and
quark-diquark picture of baryons which was previously used for the
investigation of the heavy baryon spectroscopy
\cite{hbar,hbarRegge}. In our present analysis we closely follow these considerations.  The interaction of two quarks in a diquark and the quark-diquark interaction  in a baryon are described by the
diquark wave function $\Psi_{d}$ of the bound quark-quark state
and by the baryon wave function $\Psi_{B}$ of the bound quark-diquark
state respectively,  which satisfy the
quasipotential equation of the Schr\"odinger type \cite{mass}
\begin{equation}
\label{quas}
{\left(\frac{b^2(M)}{2\mu_{R}}-\frac{{\bf
p}^2}{2\mu_{R}}\right)\Psi_{d,B}({\bf p})} =\int\frac{d^3 q}{(2\pi)^3}
 V({\bf p,q};M)\Psi_{d,B}({\bf q}),
\end{equation}
where the relativistic reduced mass is
\begin{equation}
\mu_{R}=\frac{M^4-(m^2_1-m^2_2)^2}{4M^3},
\end{equation}
and $M$ is the bound state mass (diquark or baryon),
$m_{1,2}$ are the masses of  quarks ($q_1$ and $q_2$) which form
the diquark or of the  diquark ($d$) and  quark ($q$) which form
the baryon ($B$), and ${\bf p}$  is their relative momentum.  
In the center of mass system the relative momentum squared on mass shell 
reads
\begin{equation}
{b^2(M) }
=\frac{[M^2-(m_1+m_2)^2][M^2-(m_1-m_2)^2]}{4M^2}.
\end{equation}

The kernel 
$V({\bf p,q};M)$ in Eq.~(\ref{quas}) is the quasipotential operator of
the quark-quark or quark-diquark interaction which is constructed with
the help of the
off-mass-shell scattering amplitude, projected onto the positive
energy states. We assume that the effective
interaction is the sum of the usual one-gluon exchange term and the mixture
of long-range vector and scalar linear confining potentials, where
the vector confining potential contains the Pauli term. 
The details can be found in Refs.~\cite{hbar,hbarRegge}. The resulting
quasipotentials  are given by the following expressions. 

(a) Quark-quark ($qq$) interaction in the diquark
 \begin{equation}
\label{qpot}
V({\bf p,q};M)=\bar{u}_{1}(p)\bar{u}_{2}(-p){\cal V}({\bf p}, {\bf
q};M)u_{1}(q)u_{2}(-q),
\end{equation}
with
\[
{\cal V}({\bf p,q};M)=\frac12\left[\frac43\alpha_sD_{ \mu\nu}({\bf
k})\gamma_1^{\mu}\gamma_2^{\nu}+ V^V_{\rm conf}({\bf k})
\Gamma_1^{\mu}({\bf k})\Gamma_{2;\mu}(-{\bf k})+
 V^S_{\rm conf}({\bf k})\right],
\]

(b) Quark-diquark ($qd$) interaction in the baryon
\begin{eqnarray}
\label{dpot}
V({\bf p,q};M)&=&\frac{\langle d(P)|J_{\mu}|d(Q)\rangle}
{2\sqrt{E_d(p)E_d(q)}} \bar{u}_{q}(p)  
\frac43\alpha_sD_{ \mu\nu}({\bf 
k})\gamma^{\nu}u_{q}(q)\cr
&&+\psi^*_d(P)\bar u_q(p)J_{d;\mu}\Gamma_q^\mu({\bf k})
V_{\rm conf}^V({\bf k})u_{q}(q)\psi_d(Q)\cr 
&&+\psi^*_d(P)
\bar{u}_{q}(p)V^S_{\rm conf}({\bf k})u_{q}(q)\psi_d(Q), 
\end{eqnarray}
where $\alpha_s$ is the QCD coupling constant, $\langle
d(P)|J_{\mu}|d(Q)\rangle$ is the vertex of the 
diquark-gluon interaction which takes into account the diquark
internal structure and $J_{d;\mu}$ is the effective long-range vector
vertex of the diquark. The diquark momenta are $P=(E_d(p),-{\bf p})$, $Q=(E_d(q),-{\bf q})$ with $E_d(p)=\sqrt{{\bf p}^2+M_d^2}$. $D_{\mu\nu}$ is the  
gluon propagator in the Coulomb gauge, ${\bf k=p-q}$; $\gamma_{\mu}$ and $u(p)$ are 
the Dirac matrices and spinors, while $\psi_d(P)$ is the diquark wave
function \cite{hbar}. The factor 1/2 in the quark-quark  interaction
accounts for the difference of the colour factor compared to the
quark-antiquark case.

The effective long-range vector vertex of the quark is
defined by \cite{mass}
\begin{equation}
\Gamma_{\mu}({\bf k})=\gamma_{\mu}+
\frac{i\kappa}{2m}\sigma_{\mu\nu}\tilde k^{\nu}, \qquad \tilde
k=(0,{\bf k}),
\end{equation}
where $\kappa$ is the anomalous chromomagnetic moment of quarks.

In the nonrelativistic limit the vector and scalar confining
potentials  reduce to
\begin{eqnarray}
V^V_{\rm conf}(r)&=&(1-\varepsilon)(Ar+B),\nonumber\\
V^S_{\rm conf}(r)& =&\varepsilon (Ar+B),
\end{eqnarray}
where $\varepsilon$ is the mixing coefficient. Thus in this limit the
usual Cornell-like potential is reproduced
\begin{equation}
V(r)=-\frac43\frac{\alpha_s}{r}+Ar+B,
\end{equation}
where we use the QCD coupling constant with freezing 
\begin{equation}
  \label{eq:alpha}
  \alpha_s(\mu^2)=\frac{4\pi}{\displaystyle\beta_0
\ln\frac{\mu^2+M_B^2}{\Lambda^2}}, \qquad \beta_0=11-\frac23n_f,
\qquad \mu=\frac{2m_1m_2}{m_1+m_2},
\end{equation} 
with the background mass $M_B=2.24\sqrt{A}=0.95$~GeV \cite{bvb} and
$\Lambda=413$~MeV \cite{lregge}.

 All parameters of the model were fixed previously from calculations
 of meson and baryon properties \cite{mass}. The constituent quark masses $m_u=m_d=0.33$ GeV, $m_s=0.5$ GeV  and 
the parameters of the linear potential $A=0.18$ GeV$^2$ and $B=-0.3$ GeV
have the usual values of quark models.  The value of the mixing
coefficient of vector and scalar confining potentials $\varepsilon=-1$
and the universal Pauli interaction constant $\kappa=-1$.  Note that the 
long-range chromomagnetic contribution to the potential, which
is proportional to $(1+\kappa)$,  vanishes for the 
chosen value of $\kappa=-1$.

First we calculate masses and form factors of the diquarks. The quasipotential equation (\ref{quas}) is solved numerically for the
complete relativistic potential  which depends on the
diquark mass in a complicated highly nonlinear way \cite{hbar}. In our
approach we assume that diquarks in strange baryons are formed by the
constituent quarks of the same mass, i.e. we consider only the $ud$, $uu$,
$dd$ and $ss$ diquarks. Note that the ground state $ud$ diquark can be
both in scalar and axial vector state, while the ground state diquarks
composed from quarks of the same flavour  $uu$, $dd$ and $ss$ can be
only in the axial vector state due to the Pauli principle.  The obtained
masses of the ground and excited states of diquarks are presented in
Table~\ref{tab:dmass}. The diquark state is characterized by the quark
content, isospin $I$, radial quantum number $n=1,2,3\dots$, orbital momentum $l=s,p$
and total angular momentum $j=0,1,2$ (the diquark spin).  In this table we also give the values of
the parameters $\xi$ and $\zeta$. They enter the vertex  $\langle
d(P)|J_{\mu}|d(Q)\rangle$ of the  diquark-gluon interaction
(\ref{dpot}) which is parameterized by the form factor
\begin{equation}
  \label{eq:fr}
  F(r)=1-e^{-\xi r -\zeta r^2},
\end{equation}
that takes the internal diquark structure into account \cite{hbar}.   

\begin{table}
  \caption{Masses $M$ and form factor  parameters of
   diquarks. }
  \label{tab:dmass}
\begin{ruledtabular}
\begin{tabular}{cccccc}
Quark& I&State&   $M$ &$\xi$ & $\zeta$
 \\
content& &$nl_j$ & (MeV)& (GeV)& (GeV$^2$)  \\
\hline
$ud$&0&$1s_0$ & 710 & 1.09 & 0.185  \\
&1&$1s_1$ & 909 &1.185 & 0.365  \\
&0&$1p_0$ & 1321 &1.395 & 0.148  \\
&0&$1p_1$ & 1397 &1.452 & 0.195  \\
&0&$1p_2$ & 1475 &1.595 & 0.173  \\
&1&$1p_1$ & 1392 &1.451 & 0.194  \\
&0&$2s_0$ & 1513 &1.01 & 0.055  \\
&1&$2s_1$ & 1630 &1.05 & 0.151  \\
\hline
$ss$ &0&$1s_1$ & 1203 & 1.13 & 0.280\\
 &0&$1p_1$ & 1608 & 1.03 & 0.208\\
 &0&$2s_1$ & 1817 & 0.805 & 0.235\\
  \end{tabular}
\end{ruledtabular}
\end{table}

Next we calculate the masses of heavy baryons as the bound
states of a quark and diquark. The quark-diquark  potential is the sum of spin-independent and spin-dependent parts \cite{hbar,hbarRegge}
\begin{equation}
  \label{eq:v}
  V(r)= V_{\rm SI}(r)+ V_{\rm SD}(r).
\end{equation}
The spin-independent $V_{\rm SI}(r)$ part is given by
\begin{eqnarray}
\label{si}\!\!\!\!\!\!\!
V_{\rm SI}(r)&=&\hat V_{\rm Coul}(r)
+V_{\rm  conf}(r)+\frac{1}{E_dE_{q}}\Bigg\{ \frac12(E_q^2-m_q^2+E_d^2-M_d^2)\left[\hat
  V_{\rm Coul}(r)+V^V_{\rm conf}(r)\right]\cr
&&+\frac{1}{4}\Delta \left[2V_{\rm Coul}(r)+V^V_{\rm conf}(r)\right]+\hat V'_{\rm Coul}(r)\frac{{\bf L}^2}{2r}\Bigg\}+\frac1{E_q(E_q+m_q)}\Biggl\{
-(E_q^2-m_q^2)V^S_{\rm conf}(r)\cr
&&+\frac14\Delta\left(\hat V_{\rm
Coul}(r)-V_{\rm conf}(r)-2\left[\frac{E_q-m_q}{2m_q}-(1+\kappa)\frac{E_q+m_q}{2m_q}\right]V^V_{\rm conf}(r)
\right)\Biggr\},
\end{eqnarray}
where the diquark and quark energies are defined by their
on-mass-shell values \cite{hbar}
$$E_d=\frac{M^2-m_q^2+M_d^2}{2M}, \qquad E_q=\frac{M^2-M_d^2+m_q^2}{2M}.$$ 
 Here $\Delta$ is the Laplace operator, and $\hat V_{\rm Coul}(r)$ is the smeared Coulomb
potential which accounts for the diquark internal structure
$$\hat V_{\rm Coul}(r)=-\frac43\alpha_s\frac{ F(r)}{r}.$$

The spin-dependent potential has the following form \cite{hbarRegge}
\begin{equation}
  \label{eq:vsd}
   V_{\rm SD}(r)=a_1\, {\bf L}{\bf S}_d+a_2\, {\bf L}{\bf S}_q+
b \left[-{\bf S}_d{\bf S}_q+\frac3{r^2}({\bf S}_d{\bf r})({\bf
    S}_q{\bf r})\right]+ c\, {\bf S}_d{\bf S}_q,
\end{equation}
where $ {\bf L}$ is the orbital angular momentum; ${\bf S}_d$ and
${\bf S}_q$ are the diquark and quark spin operators,
respectively. The first two terms are the  spin-orbit interactions, the
third one is the tensor interaction and the last one is the spin-spin interaction.
The coefficients $a_1$, $a_2$, $b$ and $c$ are expressed
through the corresponding derivatives of the smeared Coulomb and
confining potentials:
\begin{eqnarray}
\label{a1}
 a_1&= &\frac{1}{M_d(E_d+M_d)}\frac1{r}\Biggl[\frac{M_d}{E_d}
\hat  V'_{\rm Coul}(r)-V'_{\rm
  conf}(r)
\Biggr]\cr \cr 
&& +\frac1{E_dE_q}\frac1{r}\Biggl[
\hat V'_{\rm Coul}(r)
+\frac{E_d}{M_d}\left(\frac{E_d-M_d}{E_q+m_q}
+\frac{E_q-m_q}{E_d+M_d}\right) V'^S_{\rm conf}(r) 
\Biggr],\ \ \ \ \  \\ \cr 
\label{a2}
a_2&=&\frac{1}{E_dE_{q}}\frac1{r}\Biggl\{\hat V'_{\rm Coul}(r)
-\left[\frac{E_q-m_q}{2m_q}- 
 (1+\kappa)\frac{E_q+m_q}{2m_q}\right]V'^V_{\rm conf}(r)\Biggr\} \cr \cr 
&&+\frac{1}{E_q(E_q+m_q)}\frac1{r}\Biggl\{\hat V'_{\rm Coul}(r)-
V'_{\rm conf}(r)
-2\left[\frac{E_q-m_q}{2m_q}- 
 (1+\kappa)\frac{E_q+m_q}{2m_q}\right]V'^V_{\rm conf}(r)\Biggr\},\cr
&& \\
\label{b}
b&=&\frac13\frac{1}{E_dE_{q}}\Biggl\{\frac1{r}\hat V'_{\rm Coul}(r)-\hat V''_{\rm
    Coul}(r)
\Biggr\}, \\ \cr 
\label{c}
c&=&\frac23\frac{1}{E_dE_{q}}
\Delta \hat V_{\rm Coul}(r).
\end{eqnarray}
Note that both the one-gluon exchange and confining potentials contribute
to the quark-diquark spin-orbit interaction. The presence of the spin-orbit
${\bf L}{\bf S}_q$ and of the tensor terms in the quark-diquark potential 
(\ref{a1})--(\ref{b}) leads to a mixing of states with
the same total angular momentum $J$ and parity $P$ but different 
diquark  angular momentum  (${\bf L}+{\bf S}_d$). We consider
such mixing  in the same way as in the case of doubly heavy baryons \cite{dhbar}.

\section{Strange baryon masses}
\label{sec:sbmass}

We solve numerically the quasipotential equation with the
nonperturbative account for the
relativistic dynamics both of quarks and diquarks.   
The calculated values of the ground and excited state baryon masses are presented in
Tables~\ref{lambdapl}--\ref{omega} in comparison with available
experimental data \cite{pdg}.
In the first column we show the baryon total spin $J$ and parity 
$P$. In the next three columns experimental candidates are listed
with their status and measured mass. In the fifth column we give the
states of the quark-diquark system in a baryon and the quark-quark
state in a
diquark for which the following notations are used: $NLnl_j$, where we first show the radial quantum
number in the quark-diquark bound system ($N=1,2,3\dots$) and its orbital momentum by a
capital letter ($L=S,P,D\dots$), then the radial quantum number of two
quarks in a diquark  ($n=1,2,3\dots$), their orbital momentum by a
lowercase letter ($l=s,p,d\dots$) and their total momentum $j$ (the
diquark spin) in the
subscript. Finally, in the last column our predictions for baryon
masses are presented.

\begin{table}
\caption{Masses of the positive-parity $\Lambda$ states (in MeV).}
\label{lambdapl}
\begin{ruledtabular}
\begin{tabular}{cccccc}
&\multicolumn{3}{c}{Experiment \cite{pdg}}& \multicolumn{2}{c}{Theory}\\
\cline{2-4} \cline{5-6}
$J^P$&State&Status& Mass& $NLnl_j$& Mass\\
\hline
$\frac12^+$& $\Lambda$ & {****} & $1115.683\pm0.006$& $1S1s_0$& 1115\\
& $\Lambda(1600)$ & {***} & $1560-1600$& $2S1s_0$& 1615\\
& $\Lambda(1710)$ &*&$1713\pm13$&\\
& $\Lambda(1810)$ &  {***} & $1750-1810$& $1P1p_1$& 1901\\
&& & & $1S2s_0$& 1972\\
&& & & $1P1p_0$& 1986\\
&& & & $1P1p_2$& 2042\\
&& & & $3S1s_0$& 2099\\
&& & & $1P1p_1$& 2205\\
&& & & $2P1p_0$& 2431\\
&& & & $2S2s_0$& 2433\\
&& & & $4S1s_0$& 2546\\
&& & & $2P1p_1$& 2559\\
&& & & $2P1p_2$& 2657\\
&& & & $2P1p_1$& 2687\\
\hline
$\frac32^+$& $\Lambda(1890) $& {****} & $1850-1890$& $1D1s_0$& 1854\\
&& & & $1P1p_2$& 1976\\
&& & & $1P1p_0$& 2130\\
&& & & $1P1p_1$& 2184\\
&& & & $1P1p_2$& 2202\\
&& & & $1P1p_1$& 2212\\
&& & & $2D1s_0$& 2289\\
&& & & $2P1p_0$& 2623\\
&& & & $2P1p_2$& 2629\\
&& & & $2P1p_1$& 2690\\
&& & & $2P1p_1$& 2697\\
&& & & $2P1p_2$& 2701\\
\hline
$\frac52^+$& $\Lambda(1820) $& {****} & $1815-1820$& $1D1s_0$& 1825\\
& $\Lambda(2110) $& {***} & $2090-2110$ & $1P1p_2$& 2098\\
&& & & $1P1p_2$& 2221\\
&& & & $1P1p_1$& 2255\\
&& & & $2D1s_0$& 2258\\
&& & & $2P1p_2$& 2683\\
&& & & $2P1p_2$& 2724\\
&& & & $2P1p_1$& 2746\\
\hline
$\frac72^+$& $\Lambda(2020) $& {*} & $\approx2020$& $1P1p_2$& 2251\\
&& & & $1G1s_0$& 2471\\
&& & & $1F1p_0$& 2626\\
&& & & $2P1p_2$& 2744\\
\hline
$\frac92^+$& $\Lambda(2350) $& {***} & $2340-2350$& $1G1s_0$& 2360\\
\end{tabular}
\end{ruledtabular}
\end{table}

\begin{table}
\caption{Masses of the negative-parity $\Lambda$ states (in MeV).}
\label{lambdamin}
\begin{ruledtabular}
\begin{tabular}{cccccc}
&\multicolumn{3}{c}{Experiment \cite{pdg}}& \multicolumn{2}{c}{Theory}\\
\cline{2-4} \cline{5-6}
$J^P$&State&Status& Mass& $NLnl_j$& Mass\\
\hline
$\frac12^-$& $\Lambda(1405)$ & {****} & $1405.1^{+1.3}_{-1.0}$& $1P1s_0$& 1406\\
& $\Lambda(1670)$ & {****} & $1660-1670$& $1S1p_1$& 1667\\
& $\Lambda(1800)$ & {***} & $1720-1800$& $1S1p_0$& 1733\\
&& & & $2P1s_0$& 1927\\
&& & & $2S1p_0$& 2197\\
&& & & $1P2s_0$& 2218\\
&& & & $3P1s_0$& 2274\\
&& & & $2S1p_1$& 2290\\
&& & & $1D1p_1$& 2427\\
&& & & $1D1p_2$& 2491\\
&& & & $3S1p_0$& 2707\\
\hline
$\frac32^-$& $\Lambda(1520) $& {****} & $1519.5\pm1.0$& $1P1s_0$&
1549\\
& $\Lambda(1690) $& {****} & $1685-1690$& $1S1p_2$& 1693\\
&& & & $1S1p_1$& 1812\\
& $\Lambda(2050) $& {*} & $2056\pm22$& $2P1s_0$& 2035\\
&& & & $1P2s_0$& 2319\\
& $\Lambda(2325) $& {*} & $\approx2325$& $2S1p_2$& 2322\\
&& & & $2S1p_1$& 2392\\
&& & & $3P1s_0$& 2454\\
&& & & $1D1p_0$& 2468\\
&& & & $1D1p_1$& 2523\\
&& & & $1D1p_1$& 2546\\
&& & & $1D1p_2$& 2594\\
&& & & $1D1p_2$& 2622\\
\hline
$\frac52^-$& $\Lambda(1830) $& {****} & $1810-1830$& $1S1p_2$& 1861\\
&& & & $1F1s_0$& 2136\\
&& & & $1D1p_0$& 2350\\
&& & & $2S1p_2$& 2441\\
&& & & $1D1p_1$& 2549\\
&& & & $1D1p_1$& 2560\\
&& & & $1D1p_2$& 2625\\
&& & & $1D1p_2$& 2639\\
\hline
$\frac72^-$& $\Lambda(2100) $& {****} & $2090-2100$& $1F1s_0$& 2097\\
&& & & $1D1p_1$& 2583\\
&& & & $1D1p_2$& 2625\\
&& & & $1D1p_2$& 2639\\
\hline
$\frac92^-$& && & $1D1p_2$& 2665\\
& && & $1H1s_0$& 2738\\
\hline
$\frac{11}2^-$
& && & $1H1s_0$& 2605\\
\end{tabular}
\end{ruledtabular}
\end{table}

\begin{table}
\caption{Masses of the positive-parity $\Sigma$ states (in MeV).}
\label{sigmapl}
\begin{ruledtabular}
\begin{tabular}{cccccc}
&\multicolumn{3}{c}{Experiment \cite{pdg}}& \multicolumn{2}{c}{Theory}\\
\cline{2-4} \cline{5-6}
$J^P$&State&Status& Mass& $NLnl_j$& Mass\\
\hline
$\frac12^+$& $\Sigma$ & {****} & $1189.37\pm0.07$& $1S1s_1$& 1187\\
& $\Sigma(1660)$ & {***} & $1630-1660$& $2S1s_1$& 1711\\
& $\Sigma(1770)$ &  {*} & $\approx1770$& $1P1p_1$& 1922\\
& $\Sigma(1880)$ &  {*} & $\approx1880$& $1D1s_1$& 1983\\
&& & & $1S2s_1$& 2028\\
&& & & $1P1p_1$& 2180\\
&& & & $3S1s_1$& 2292\\
&& & & $2D1s_1$& 2472\\
&& & & $2P1p_1$& 2515\\
&& & & $2S2s_1$& 2530\\
&& & & $2P1p_1$& 2647\\
&& & & $1D2s_1$& 2672\\
&& & & $4S1s_1$& 2740\\
\hline
$\frac32^+$& $\Sigma(1385) $& {****} & $1382.80\pm0.35$& $1S1s_1$&
1381\\
& $\Sigma(1730) $& {*} & $1727\pm27$&\\
& $\Sigma(1840) $& {*} & $\approx1840$& $2S1s_1$& 1862\\
& $\Sigma(1940) $& {*} & $1941\pm18$& $1D1s_1$& 2025\\
& $\Sigma(2080) $& {**} & $\approx2080$& $1D1s_1$& 2076\\
&& & & $1S2s_1$& 2096\\
&& & & $1P1p_1$& 2157\\
&& & & $1P1p_1$& 2186\\
&& & & $3S1s_1$& 2347\\
&& & & $2D1s_1$& 2465\\
&& & & $2D1s_1$& 2483\\
&& & & $2S2s_1$& 2584\\
&& & & $2P1p_1$& 2640\\
&& & & $2P1p_1$& 2654\\
\hline
$\frac52^+$& $\Sigma(1915) $& {****} & $1900-1915$& $1D1s_1$& 1991\\
& $\Sigma(2070) $& {*} & $\approx2070$ & $1D1s_1$& 2062\\
&& & & $1P1p_1$& 2221\\
&& & & $2D1s_1$& 2459\\
&& & & $2D1s_1$& 2485\\
&& & & $2P1p_1$& 2701\\
\hline
$\frac72^+$& $\Sigma(2030) $& {****} & $2025-2030$& $1D1s_1$& 2033\\
&& & & $2D1s_1$& 2470\\
&& & & $1G1s_1$& 2619\\
\hline
$\frac92^+$& & && $1G1s_1$& 2548\\
& & && $1G1s_1$& 2619\\
\hline
$\frac{11}2^+$& & && $1G1s_1$& 2529\\
\end{tabular}
\end{ruledtabular}
\end{table}

\begin{table}
\caption{Masses of the negative-parity $\Sigma$ states (in MeV).}
\label{sigmamin}
\begin{ruledtabular}
\begin{tabular}{cccccc}
&\multicolumn{3}{c}{Experiment \cite{pdg}}& \multicolumn{2}{c}{Theory}\\
\cline{2-4} \cline{5-6}
$J^P$&State&Status& Mass& $NLnl_j$& Mass\\
\hline
$\frac12^-$& $\Sigma(1620)$ & {*} & $\approx1620$& $1P1s_1$& 1620\\
&& & & $1S1p_1$& 1693\\
& $\Sigma(1750)$ & {***} & $1730-1750$& $1P1s_1$& 1747\\
& $\Sigma(1900)$ & {*} & $1900\pm21$& $2P1s_1$& 2115\\
& $\Sigma(2000)$ &  {*} & $\approx2000$& $2P1s_1$& 2198\\
&& & & $2S1p_1$& 2202\\
&& & & $1P2s_1$& 2289\\
&& & & $1D1p_1$& 2381\\
&& & & $1P2s_1$& 2427\\
&& & & $3P1s_1$& 2630\\
&& & & $3P1s_1$& 2634\\
&& & & $3S1p_1$& 2742\\
\hline
$\frac32^-$& $\Sigma(1580) $& {*} & $\approx1580$& \\
& $\Sigma(1670) $& {***} & $1665-1670$&$1P1s_1$& 1706\\
&& && $1P1s_1$& 1731\\
& $\Sigma(1940) $& {***} & $1900-1940$& $1S1p_1$& 1856\\
& && & $2P1s_1$& 2175\\
&& & & $2P1s_1$& 2203\\
&& & & $2S1p_1$& 2300\\
&& & & $1F1s_1$& 2409\\
&& & & $1P2s_1$& 2410\\
&& & & $1P2s_1$& 2430\\
&& & & $1D1p_1$& 2494\\
&& & & $1D1p_1$& 2513\\
&& & & $3P1s_1$& 2623\\
&& & & $3P1s_1$& 2637\\
\hline
$\frac52^-$& $\Sigma(1775) $& {****} & $1770-1775$& $1P1s_1$& 1757\\
& && & $2P1s_1$& 2214\\
&& & & $1F1s_1$& 2347\\
&& & & $1P2s_1$& 2459\\
&& & & $1F1s_1$& 2475\\
&& & & $1D1p_1$& 2516\\
&& & & $1D1p_1$& 2524\\
& && & $3P1s_1$& 2644\\
\hline
$\frac72^-$& $\Sigma(2100) $& {*} & $\approx2100$& $1F1s_1$& 2259\\
&& & & $1F1s_1$& 2349\\
&& & & $1D1p_1$& 2545\\
\hline
$\frac92^-$& & && $1F1s_1$& 2289\\
\end{tabular}
\end{ruledtabular}
\end{table}

\begin{table}
\caption{Masses of the positive-parity $\Xi$ states (in MeV).}
\label{xipl}
\begin{ruledtabular}
\begin{tabular}{cccccc}
&\multicolumn{3}{c}{Experiment \cite{pdg}}& \multicolumn{2}{c}{Theory}\\
\cline{2-4} \cline{5-6}
$J^P$&State&Status& Mass& $NLnl_j$& Mass\\
\hline
$\frac12^+$& $\Xi$ & {****} & $1321.71\pm0.07$& $1S1s_1$& 1330\\
& &  && $2S1s_1$& 1886\\
& & & & $1D1s_1$& 1993\\
&  & & & $1P1p_1$& 2012\\
&& & & $1S2s_1$& 2091\\
&& & & $1P1p_1$& 2142\\
&& & & $3S1s_1$& 2367\\
&& & & $2S2s_1$& 2456\\
&& & & $2D1s_1$& 2510\\
&& & & $1D2s_1$& 2565\\
&  & & & $2P1p_1$& 2598\\
&  & & & $2P1p_1$& 2624\\
\hline
$\frac32^+$& $\Xi(1530) $& {****} & $1531.80\pm0.32$& $1S1s_1$&
1518\\
& & && $2S1s_1$& 1966\\
& & & & $1D1s_1$& 2100\\
&& & & $1S2s_1$& 2121\\
& & & & $1D1s_1$& 2122\\
&& & & $1P1p_1$& 2144\\
&& & & $1P1p_1$& 2149\\
&& & & $3S1s_1$& 2421\\
&& & & $2S2s_1$& 2491\\
&& & & $2D1s_1$& 2597\\
&  & & & $2P1p_1$& 2640\\
&& & & $2D1s_1$& 2663\\
&  & & & $2P1p_1$& 2664\\
\hline
$\frac52^+$& & && $1D1s_1$& 2108\\
& & & & $1D1s_1$& 2147\\
&& & & $1P1p_1$& 2213\\
&& & & $2D1s_1$& 2605\\
&& & & $2D1s_1$& 2630\\
\hline
$\frac72^+$&& && $1D1s_1$& 2189\\
&& & & $2D1s_1$& 2686\\
\end{tabular}
\end{ruledtabular}
\end{table}

\begin{table}
\caption{Masses of the negative-parity $\Xi$ states (in MeV).}
\label{ximin}
\begin{ruledtabular}
\begin{tabular}{cccccc}
&\multicolumn{3}{c}{Experiment \cite{pdg}}& \multicolumn{2}{c}{Theory}\\
\cline{2-4} \cline{5-6}
$J^P$&State&Status& Mass&$NLnl_j$& Mass\\
\hline
$\frac12^-$&  & && $1P1s_1$& 1682\\
& &  && $1P1s_1$& 1758\\
& & & & $1S1p_1$& 1839\\
&  & & & $2P1s_1$& 2160\\
&& & & $2S1p_1$& 2210\\
&& & & $2P1s_1$& 2233\\
&& & & $1P2s_1$& 2261\\
&& & & $1D1p_1$& 2346\\
&& & & $1P2s_1$& 2347\\
\hline
$\frac32^-$&& & & $1P1s_1$&
1764\\
&$\Xi(1820) $ & {***} & $1823\pm5$& $1P1s_1$& 1798\\
& & & & $1S1p_1$& 1904\\
&& & & $2P1s_1$& 2245\\
& & & & $2P1s_1$& 2252\\
&& & & $1P2s_1$& 2350\\
&& & & $1P2s_1$& 2352\\
&& & & $1F1s_1$& 2400\\
&& & & $1D1p_1$& 2482\\
&& & & $1D1p_1$& 2506\\
\hline
$\frac52^-$& & && $1P1s_1$& 1853\\
& & && $2P1s_1$& 2333\\
& & && $1P2s_1$& 2411\\
& & & & $1F1s_1$& 2455\\
&& & & $1D1p_1$& 2489\\
&& & & $1D1p_1$& 2545\\
&& & & $1F1s_1$& 2569\\
\hline
$\frac72^-$&& && $1F1s_1$& 2460\\
&& && $1F1s_1$& 2474\\
&& & & $1D1p_1$& 2611\\
\hline
$\frac92^-$&& && $1F1s_1$& 2502\\
\end{tabular}
\end{ruledtabular}
\end{table}

\begin{table}
\caption{Masses of the  $\Omega$ states (in MeV).}
\label{omega}
\begin{ruledtabular}
\begin{tabular}{cccccc@{\ \ \ }||ccc}
&\multicolumn{3}{c}{Experiment \cite{pdg}}&
\multicolumn{2}{c||}{Theory}& & \multicolumn{2}{c}{Theory}\\
\cline{2-4} \cline{5-6}\cline{8-9}
$J^P$&State&Status& Mass&$NLnl_j$ & Mass&$J^P$&$NLnl_j$& Mass\\
\hline
$\frac12^+$&  & && $1D1s_1$& 2301&$\frac12^-$& $1P1s_1$& 1941 \\
&&&&&& & $2P1s_1$& 2463\\
&&&&&& & $1P2s_1$& 2580\\
\hline
$\frac32^+$&$\Omega$& {****} & $1672.45\pm0.29$ & $1S1s_1$&
1678& $\frac32^-$&$1P1s_1$&2038\\
& & & & $2S1s_1$& 2173&&$2P1s_1$&2537\\
& & & & $1S2s_1$& 2304& &$1P2s_1$&2636\\
&& & & $1D1s_1$& 2332& &\\
& & & & $3S1s_1$& 2671& &\\
\hline
$\frac52^+$& & && $1D1s_1$& 2401&$\frac52^-$& $1F1s_1$& 2653\\
\hline
$\frac72^+$&& && $1D1s_1$& 2369&$\frac72^-$& $1F1s_1$& 2599\\
\hline
&&&&&& $\frac92^-$& $1F1s_1$& 2649\\
\end{tabular}
\end{ruledtabular}
\end{table}

From Tables~\ref{lambdapl}--\ref{omega} we see that  most of the
observed 3- and 4-star states of strange baryons can be well described as
ground and excited states of the quark-diquark bound system in which
diquark is in the ground either scalar or axial vector state. However
not all of these experimental states can be reproduced. Main deviations from this
picture are found in the $\Lambda$ 
sector which is better studied experimentally. Indeed the observed
$\frac12^-$  4-star states
$\Lambda(1405)$ and $\Lambda(1670)$;  $\frac32^-$  4-star states
$\Lambda(1520)$ and $\Lambda(1690)$; $\frac12^+$  3-star states
$\Lambda(1600)$ and $\Lambda(1810)$ as well as   $\frac52^+$ 4-star
$\Lambda(1820)$ and  3-star $\Lambda(2110)$ cannot be simultaneously
described in such a simple picture since their mass differences (about 200~MeV) are too
small to be attributed to the radial excitations in the quark-diquark bound system
amounting to about 500~MeV. Therefore the consideration of excitations
inside diquarks is necessary. As we can see from
Tables~\ref{lambdapl}--\ref{omega} the account of diquark excitations
allows us to  describe all these states and, as a result, to get good
agreement of the obtained predictions with data. 

In Tables~\ref{lambdacomp}--\ref{omegacomp} we compare the results of
our model with previous predictions in various theoretical
approaches. The strange baryons were treated in a relativized version
of the quark potential model in Ref.~\cite{ci}. The relativistically
covariant quark model based on the Bethe-Salpeter equation with
instantaneous two- and three-body forces was employed in
Ref.~\cite{lmp}. In Ref.~ \cite{mps} the  relativistic
quark model with the interquark interaction  arising from the meson
exchange was used.  The authors of Ref.~\cite{sf} made their
calculations of baryon masses below 2 GeV in the relativistic
interacting quark-diquark model with 
the G\"ursey and Radicati-inspired exchange interaction. Note that in
contrast to our approach, all possible types of ground-state scalar
and axial vector diquarks, including $qs$ ($q=u$ or $d$), were used in
Ref.~\cite{sf}, but excitations of diquarks were not
considered. Finally, the results of lattice calculations with two light
dynamical chirally improved quarks corresponding to pion masses between
255 and 596 MeV \cite{elms} are given.

\begin{table}
\vspace*{-0.5cm}\caption{Comparison of theoretical predictions and experimental data for the masses of the $\Lambda$ states (in MeV).}
\label{lambdacomp}
\begin{ruledtabular}
\begin{tabular}{cccccccccc}
&\multicolumn{3}{c}{Experiment \cite{pdg}}& \multicolumn{6}{c}{Theory}\\
\cline{2-4} \cline{5-10}
$J^P$&State&Status& Mass&Our & \cite{ci}&\cite{lmp}&\cite{mps}&\cite{sf}&\cite{elms}\\
\hline
$\frac12^+$& $\Lambda$ & {****} & $1115.683\pm0.006$& 1115&1115&1108&1136&1116&$1149\pm18$\\
& $\Lambda(1600)$ & {***} & $1560-1600$&
1615&1680&1677&1625&1518&$1807\pm94$\\
& $\Lambda(1710)$ &*&$1713\pm13$&\\
& $\Lambda(1810)$ &  {***} & $1750-1810$& 1901&1830&1747&1799&1666&$2112\pm54$\\
&& & & 1972&1910&1898&&1955&$2137\pm69$\\
&& & & 1986&2010&2077&&1960\\
&& & & 2042&2105&2099\\
&& & & 2099&2120&2132\\
\hline
$\frac32^+$& $\Lambda(1890) $& {****} & $1850-1890$& 1854&1900&1823&&1896&$1991\pm103$\\
&& & & 1976&1960&1952&&&$2058\pm139$\\
&& & & 2130&1995& 2045 &&&$2481\pm111$\\
&& & & 2184&2050&2087\\
&& & & 2202&2080&2133\\
\hline
$\frac52^+$& $\Lambda(1820) $& {****} & $1815-1820$& 1825&1890&1834&
&1896\\
& $\Lambda(2110) $& {***} & $2090-2110$ & 2098&2035&1999\\
&& & &  2221&2115&2078\\
&& & &  2255&2115&2127\\
&& & & 2258 &2180&2150\\
\hline
$\frac72^+$& $\Lambda(2020) $& {*} & $\approx2020$&  2251&2120&2130\\
&& & & 2471& &2331\\
\hline
$\frac92^+$& $\Lambda(2350) $& {***} & $2340-2350$&  2360& & 2340\\
\hline
$\frac12^-$& $\Lambda(1405)$ & {****} & $1405.1^{+1.3}_{-1.0}$& 1406&1550&1524&1556&1431&$1416\pm81$\\
& $\Lambda(1670)$ & {****} & $1660-1670$& 1667&1615&1630&1682&1443&$1546\pm110$\\
& $\Lambda(1800)$ &  {***} & $1720-1800$& 1733&1675&1816&1778&1650&$1713\pm116$\\
&& & &  1927&2015& 2011&&1732&$2075\pm249$\\
&& & &  2197&2095&2076&&1785\\
&& & & 2218&2160&2117& &1854\\
\hline
$\frac32^-$& $\Lambda(1520) $& {****} & $1519.5\pm1.0$& 
1549&1545&1508&1556&1431&$1751\pm40$\\
& $\Lambda(1690) $& {****} & $1685-1690$&  1693&1645&1662&1682&1443&$2203\pm106$\\
&& & & 1812&1770&1775&&1650&$2381\pm87$\\
& $\Lambda(2050) $& {*} & $2056\pm22$&  2035&2030&1987&&1732\\
& & && 2319&2110&2090&&1785\\
&$\Lambda(2325) $& {*} & $\approx2325$ &  2322&2185&2147&&1854\\
&& & &  2392&2230&2259&&1928\\
&& & & 2454&2290&2275&&1969\\
&& & &  2468&&2313\\
\hline
$\frac52^-$& $\Lambda(1830) $& {****} & $1810-1830$&  1861&1775&1828&1778&1785\\
&& & &  2136&2180&2080\\
&& & &  2350&2250&2179\\
\hline
$\frac72^-$& $\Lambda(2100) $& {****} & $2090-2100$& 2097&2150&2090\\
&& & & 2583&2230&2227\\
\hline
$\frac92^-$& && &  2665&&2370\\
\end{tabular}
\end{ruledtabular}
\end{table}

\begin{table}
\vspace*{-0.5cm}\caption{Comparison of theoretical predictions and experimental data for the masses of the $\Sigma$ states (in MeV).}
\label{sigmacomp}
\begin{ruledtabular}
\begin{tabular}{cccccccccc}
&\multicolumn{3}{c}{Experiment \cite{pdg}}& \multicolumn{6}{c}{Theory}\\
\cline{2-4} \cline{5-10}
$J^P$&State&Status& Mass&Our & \cite{ci}&\cite{lmp}&\cite{mps}&\cite{sf}&\cite{elms}\\
\hline
$\frac12^+$& $\Sigma$ & {****} & $1189.37\pm0.07$& 1187&1190 &1190& 1180&1211&$1216\pm15$\\
& $\Sigma(1660)$ & {***} & $1630-1660$& 1711&1720&1760&1616&1546&$2069\pm74$\\
& $\Sigma(1770)$ &  {*} & $\approx1770$& 1922&1915&1947&1911&1668&$2149\pm66$\\
& $\Sigma(1880)$ &  {*} & $\approx1880$& 1983&1970&2009&&1801&$2335\pm63$\\
&& & & 2028&2005&2052\\
&& & & 2180&2030&2098\\
&& & & 2292&2105&2138\\
&& & & 2472&2195\\
\hline
$\frac32^+$& $\Sigma(1385) $& {****} & $1382.80\pm0.35$& 
1381&1370&1411&1389&1334&$1471\pm23$\\
& $\Sigma(1730) $& {*} & $1727\pm27$&&1920&1896&1865&1439\\
& $\Sigma(1840) $& {*} & $\approx1840$&  1862&1970&1961&&1924&$2194\pm81$\\
& $\Sigma(1940) $& {*} & $1941\pm18$& 2025&2010&2011&&&$2250\pm79$\\
& $\Sigma(2080) $& {**} & $\approx2080$&2076&2030&2044&&&$2468\pm67$\\
&& & &  2096&2045&2062\\
&& & & 2157&2085&2103\\
&& & & 2186&2115&2112\\
\hline
$\frac52^+$& $\Sigma(1915) $& {****} & $1900-1915$&  1991&1995&1956&&2061\\
& $\Sigma(2070) $& {*} & $\approx2070$ &  2062&2030&2027\\
&& & & 2221&2095&2071\\
\hline
$\frac72^+$& $\Sigma(2030) $& {****} & $2025-2030$&  2033&2060&2070\\
&& & & 2470&2125&2161\\
\hline
$\frac12^-$& $\Sigma(1620)$ & {*} & $\approx1620$& 1620&1630&1628&1677&1753&$1603\pm38$\\
&& & & 1693&1675&1771&1736&1868&$1718\pm58$\\
& $\Sigma(1750)$ & {***} & $1730-1750$& 1747&1695&1798&1759&1895&$1730\pm34$\\
& $\Sigma(1900)$ &  {*} & $1900\pm21$&  2115&2110&2111&&&$2478\pm104$\\
& $\Sigma(2000)$ &  {*} & $\approx2000$&  2198&2155&2136\\
&& & &  2202&2165&2251\\
&& & &  2289&2205&2264\\
&& & &  2381&2260&2288\\
\hline
$\frac32^-$& $\Sigma(1580) $& {*} & $\approx1580$& \\
& $\Sigma(1670) $& {***} & $1665-1670$& 1706&1655&1669&1677&1753&$1736\pm40$\\
&& && 1731&1750&1728&1736&1868&$1861\pm20$\\
& $\Sigma(1940) $& {***} & $1900-1940$&  1856&1755&1781&1759&1895&$2297\pm122$\\
& && &  2175&2120&2139&&&$2394\pm74$\\
&& & &  2203&2185&2171\\
&& & &  2300&2200&2203\\
\hline
$\frac52^-$& $\Sigma(1775) $& {****} & $1770-1775$&  1757&1755&1770&1736&1753\\
& && & 2214&2205&2174\\
&& & &  2347&2250&2226\\
\hline
$\frac72^-$& $\Sigma(2100) $& {*} & $\approx2100$&  2259&2245&2236\\
&& & & 2349&&2285\\
\hline
$\frac92^-$& & &&  2289&&2325\\
\end{tabular}
\end{ruledtabular}
\end{table}

\begin{table}
\caption{Comparison of theoretical predictions and experimental data for the masses of the $\Xi$ states (in MeV).}
\label{xicomp}
\begin{ruledtabular}
\begin{tabular}{cccccccccc}
&\multicolumn{3}{c}{Experiment \cite{pdg}}& \multicolumn{6}{c}{Theory}\\
\cline{2-4} \cline{5-10}
$J^P$&State&Status& Mass&Our & \cite{ci}&\cite{lmp}&\cite{mps}&\cite{sf}&\cite{elms}\\
\hline
$\frac12^+$& $\Xi$ & {****} & $1321.71\pm0.07$& 1330&1305&1310&1348&1317&$1303\pm13$\\
& &  &&  1886&1840&1876&1805&1772&$2178\pm48$\\
& & & & 1993&2040&2062&&1868&$2231\pm44$\\
&  & & & 2012&2100&2131&&1874&$2408\pm45$\\
&& & & 2091&2130&2176\\
&& & & 2142&2150&2215\\
&& & & 2367&2230&2249\\
\hline
$\frac32^+$& $\Xi(1530) $& {****} & $1531.80\pm0.32$& 
1518&1505&1539&1528&1552&$1553\pm18$\\
& & && 1966& 2045&1988&&1653&$2228\pm44$\\
& & & & 2100&2065&2076&&&$2398\pm52$\\
&& & & 2121&2115&2128&&&$2574\pm52$\\
& & & & 2122&2165&2170\\
&& & &  2144&2170&2175\\
&& & &  2149&2210&2219\\
&& & & 2421&2230&2257\\
\hline
$\frac52^+$& & && 2108&2045&2013\\
& & & & 2147&2165&2141\\
&& & &  2213&2230&2197\\
\hline
$\frac72^+$&& && 2189&2180&2169\\
\hline
$\frac12^-$&  & && 1682&1755&1770&&&$1716\pm43$\\
& &  && 1758&1810&1922&&&$1837\pm28$\\
& & & & 1839&1835&1938&&&$1844\pm43$\\
&  & & & 2160&2225&2241&&&$2758\pm78$\\
&& & &  2210&2285&2266\\
&& & &  2233&2300&2387\\
&& & & 2261&2320&2411\\
\hline
$\frac32^-$&& & &
1764&1785&1780&1792&1861&$1894\pm38$\\
&$\Xi(1820) $ & {***} & $1823\pm5$& 1798&1880&1873&&1971&$1906\pm29$\\
& & & &  1904&1895&1924&&&$2426\pm73$\\
&& & & 2245&2240&2246&&&$2497\pm61$\\
& & & & 2252&2305&2284\\
&& & & 2350&2330&2353\\
&& & & 2352&2340&2384\\
\hline
$\frac52^-$& & && 1853&1900&1955&1881\\
& & && 2333&2345&2292\\
& & && 2411&2350&2409\\
\hline
$\frac72^-$&& && 2460&2355&2320\\
&& && 2474&&2425\\
\hline
$\frac92^-$&& && 2502&&2505\\
\end{tabular}
\end{ruledtabular}
\end{table}

\begin{table}
\caption{Comparison of theoretical predictions and experimental data for the masses of the $\Omega$ states (in MeV).}
\label{omegacomp}
\begin{ruledtabular}
\begin{tabular}{ccccccccc}
&\multicolumn{3}{c}{Experiment \cite{pdg}}& \multicolumn{5}{c}{Theory}\\
\cline{2-4} \cline{5-9}
$J^P$&State&Status& Mass&Our & \cite{ci}&\cite{lmp}&\cite{sf}&\cite{elms}\\
\hline
$\frac12^+$&  & && 2301&2220&2232&&$2350\pm63$\\
&&&&&2255&2256&&$2481\pm51$\\
\hline
$\frac32^+$&$\Omega$& {****} & $1672.45\pm0.29$ &
1678&1635&1636&1672&$1642\pm17$\\
& & & & 2173&2165&2177&&$2470\pm49$\\
& & & & 2304&2280&2236 \\
&& & &  2332&2345 &2287\\
\hline
$\frac52^+$& & &&  2401&2280&2253\\
&&&&&2345&2312\\
\hline
$\frac72^+$&& && 2369&2295&2292\\
\hline
$\frac12^-$& &&& 1941&1950&1992&&$1944\pm56$ \\
 &&&& 2463&2410&2456&&$2716\pm118$\\
& &&& 2580&2490&2498\\
\hline
 $\frac32^-$&&&&2038&2000&1976&&$2049\pm32$\\
&&&&2537&2440&2446&&$2755\pm67$\\
&&&&2636&2495&2507\\
\hline
$\frac52^-$& &&& 2653&2490&2528\\
\hline
$\frac72^-$& &&& 2599&&2531\\
\hline
$\frac92^-$& &&& 2649&&2606\\
\end{tabular}
\end{ruledtabular}
\end{table}

From these tables we see that our diquark model predicts appreciably less states
than the three-quark approaches. The differences become apparent with
the growth of the orbital and radial excitations in the baryon. Our
results turn out to be competitive with their predictions for the masses of the
well established (4- and 3-star) resonances, which agree well with
experimental data. For the less established (1- and 2-star) states
situation is more complicated.

First we discuss results for the $\Lambda$ sector. It is necessary to
emphasis that the experimental mass of the $\frac12^-$ 4-star $\Lambda(1405)$ is naturally
reproduced if this state is considered as the first orbital excitation
$1P$ in the strange quark-light scalar ($1s_0$) diquark 
picture of $\Lambda$  baryons. The rather low mass of this state  represents
difficulties for most of the three-quark models \cite{ci,lmp,mps},
which predict its mass about 100~MeV higher than experimental value.
There are no theoretical candidates for the $\frac12^+$ 1-star
$\Lambda(1710)$ state.
The mass of the $\frac72^+$ 1-star $\Lambda(2020)$ state is predicted
somewhat heavier by all models. Other 1-star $\Lambda$ states are well
described. 

In the $\Sigma$ sector all considered approaches cannot accommodate the $\frac32^-$
1-star $\Sigma(1580)$ state.  The predicted lowest mass $\frac32^-$
state corresponds to the 3-star  $\Sigma(1670)$ state. We have no
candidate for the $\frac32^+$ 1-star $\Sigma(1730)$ state in our
model. The calculated masses of the 1-star $\frac12^+$
$\Sigma(1770)$, $\frac12^-$ $\Sigma(1900)$ and $\frac72^-$
$\Sigma(2100)$ candidates are by more than 100~MeV heavier than experimentally
measured masses. All other known 2- and 1-star $\Sigma$  states are described
with reasonable accuracy.  

In the $\Xi$ sector only three  (two 4- and one 3-star) states and in the $\Omega$
sector only one (4-star) state  of
the observed  baryons have established quantum numbers. They are well
described by our model. We have at
least one candidate for each of the other eight $\Xi$ (three of them
have 3-stars) and three $\Omega$ (one of them has 3-stars)
states given in PDG Listings \cite{pdg} with the
predicted masses close to the experimental values. However it will be
too speculative to assign the quantum numbers to these states only on
the basis of their masses. More experimental and theoretical input is
needed. 

\section{Regge trajectories of strange baryons}
\label{sec:rtsb}

In the presented analysis we calculated masses of orbitally excited strange baryons up to rather high orbital excitation
numbers: up to $L=5$ in the quark-diquark bound system, where diquark is in the
ground state. This makes it possible 
to construct the strange baryon Regge trajectories:  
\begin{equation}
  \label{eq:reggej}
J=\alpha M^2+\alpha_0,
\end{equation}
where $\alpha$ is the slope and  $\alpha_0$ is the intercept.

\begin{figure}[htb]
 \includegraphics[width=8cm]{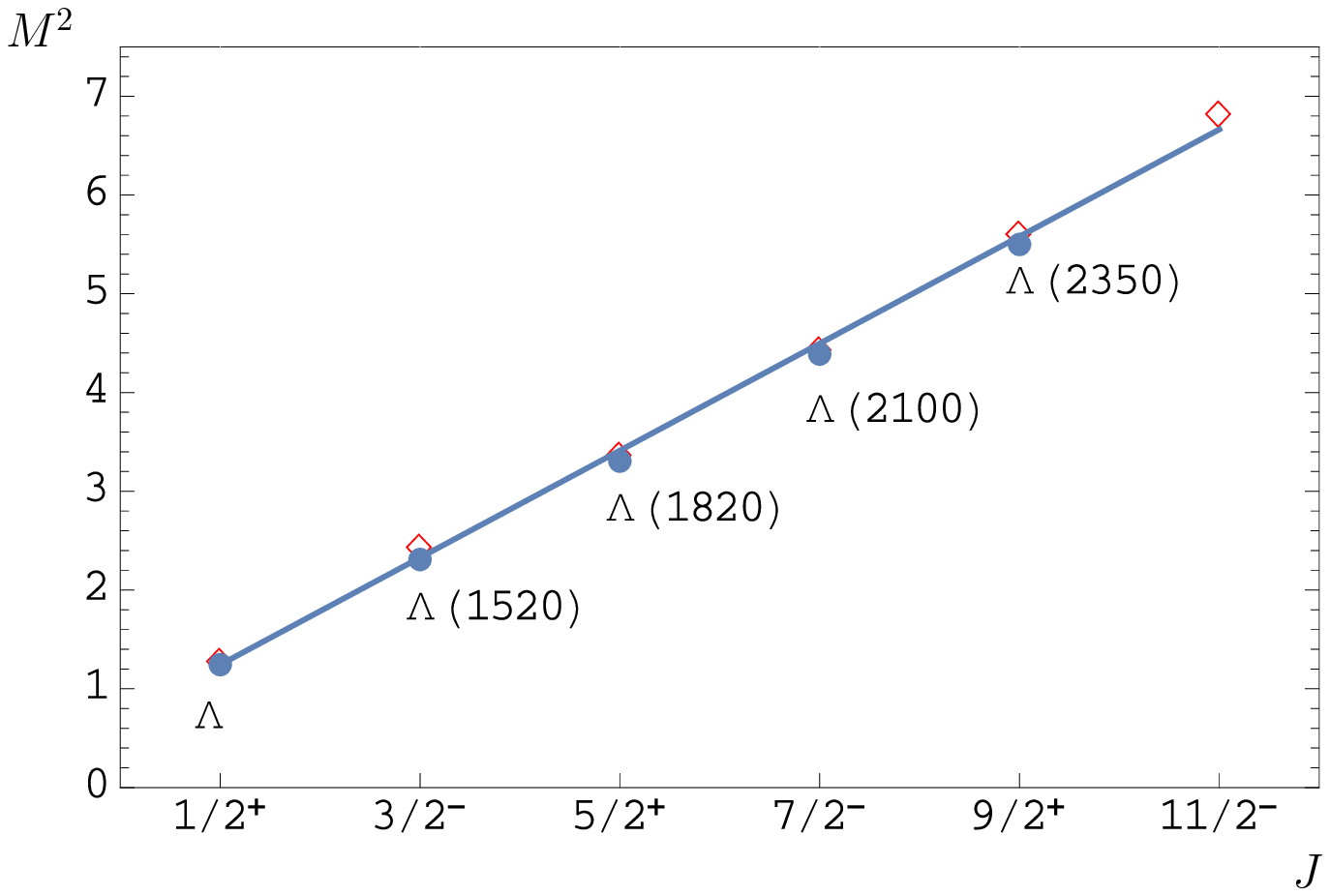}\ \ \ \includegraphics[width=8cm]{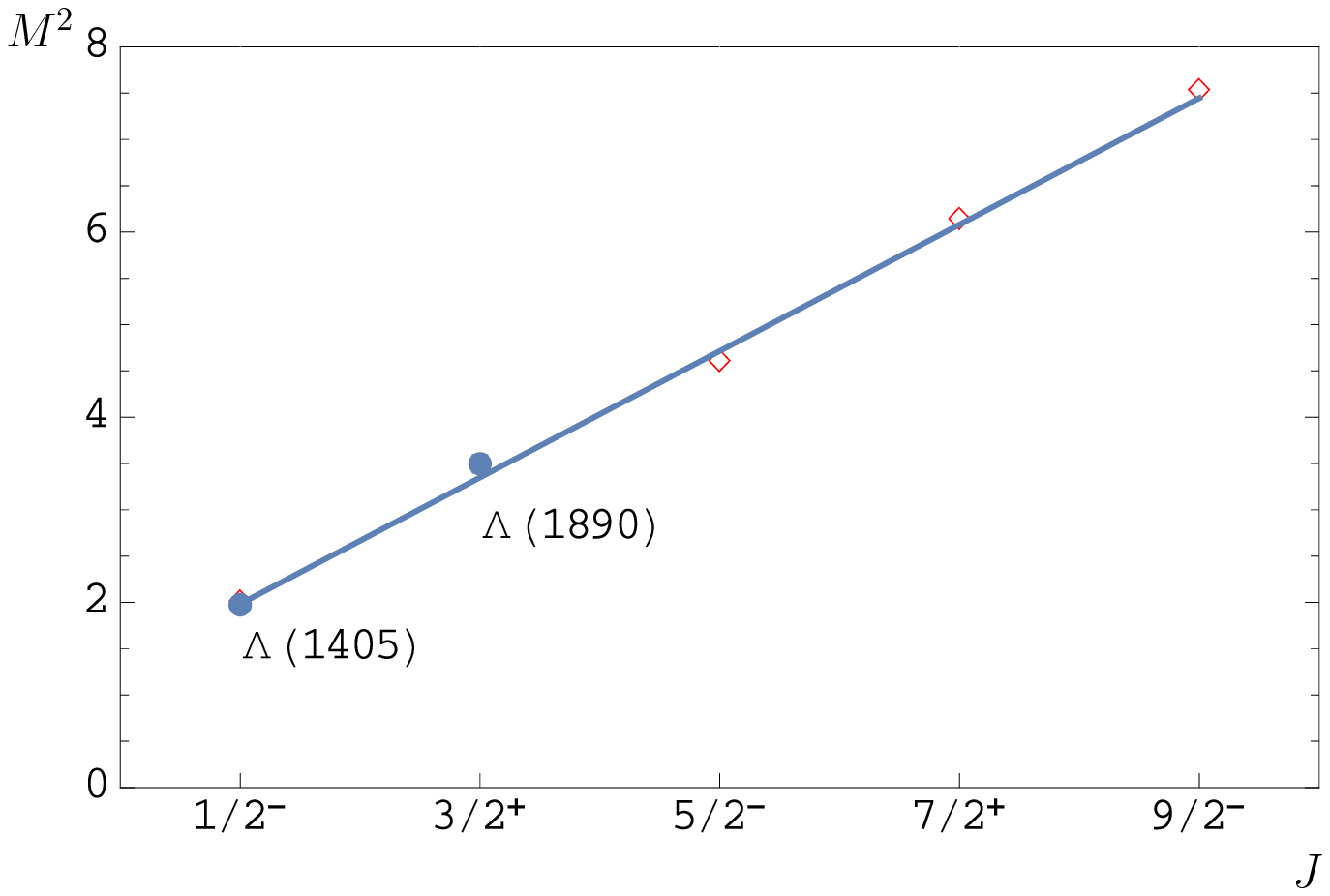}\\
\vspace*{-0.3cm}\hspace*{0.5cm} (a) \hspace*{7.5cm} (b) 
\caption{\label{fig:lambda_j} The ($J, M^2$) Regge trajectories for
  the $\Lambda$ baryons with natural (a) and unnatural (b) parities. Diamonds are predicted
  masses. Available experimental data are given by dots with  particle
  names; $M^2$ is in GeV$^2$. } 
\end{figure}

\begin{figure}
 \includegraphics[width=8.cm]{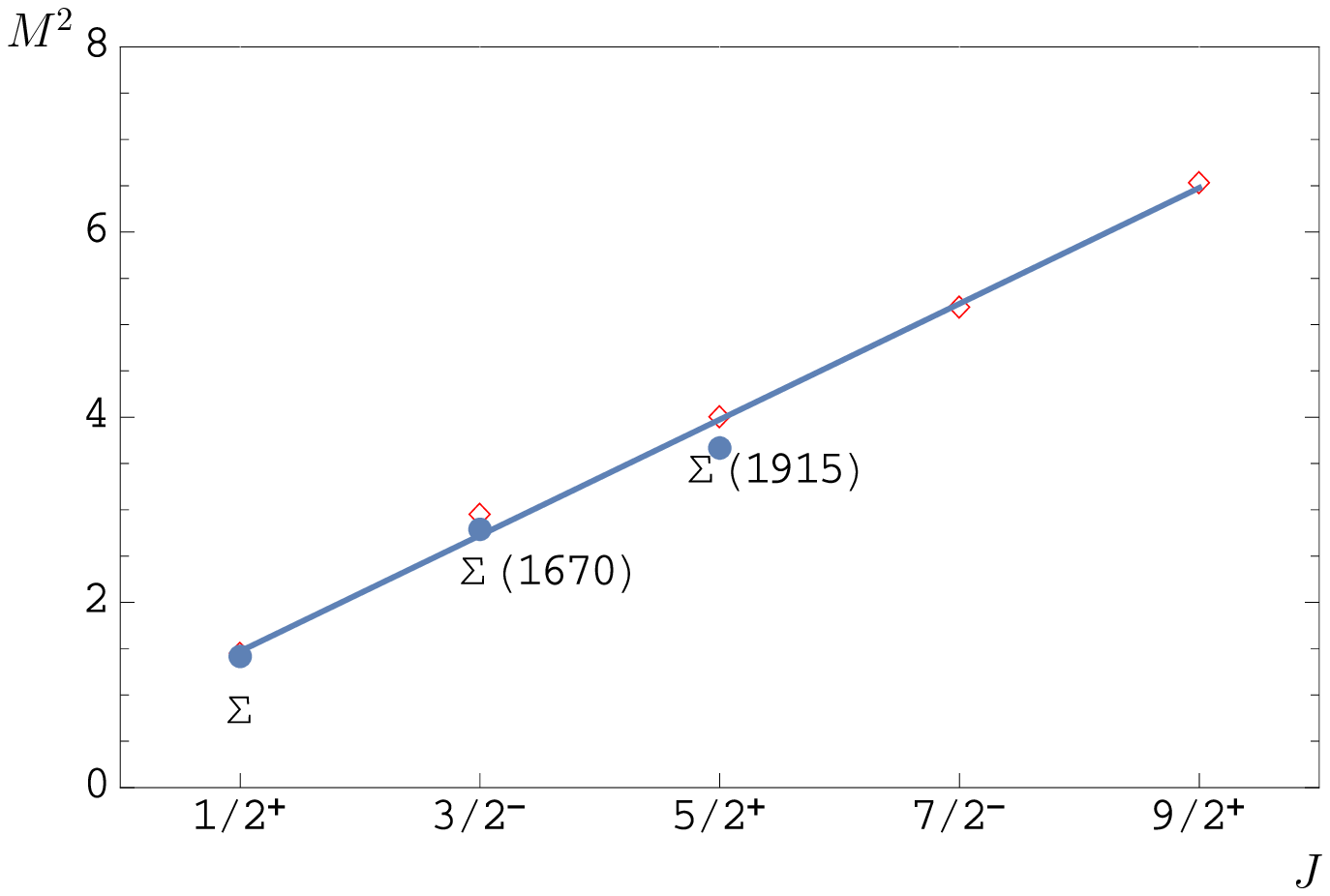}\ \
 \ \includegraphics[width=8.cm]{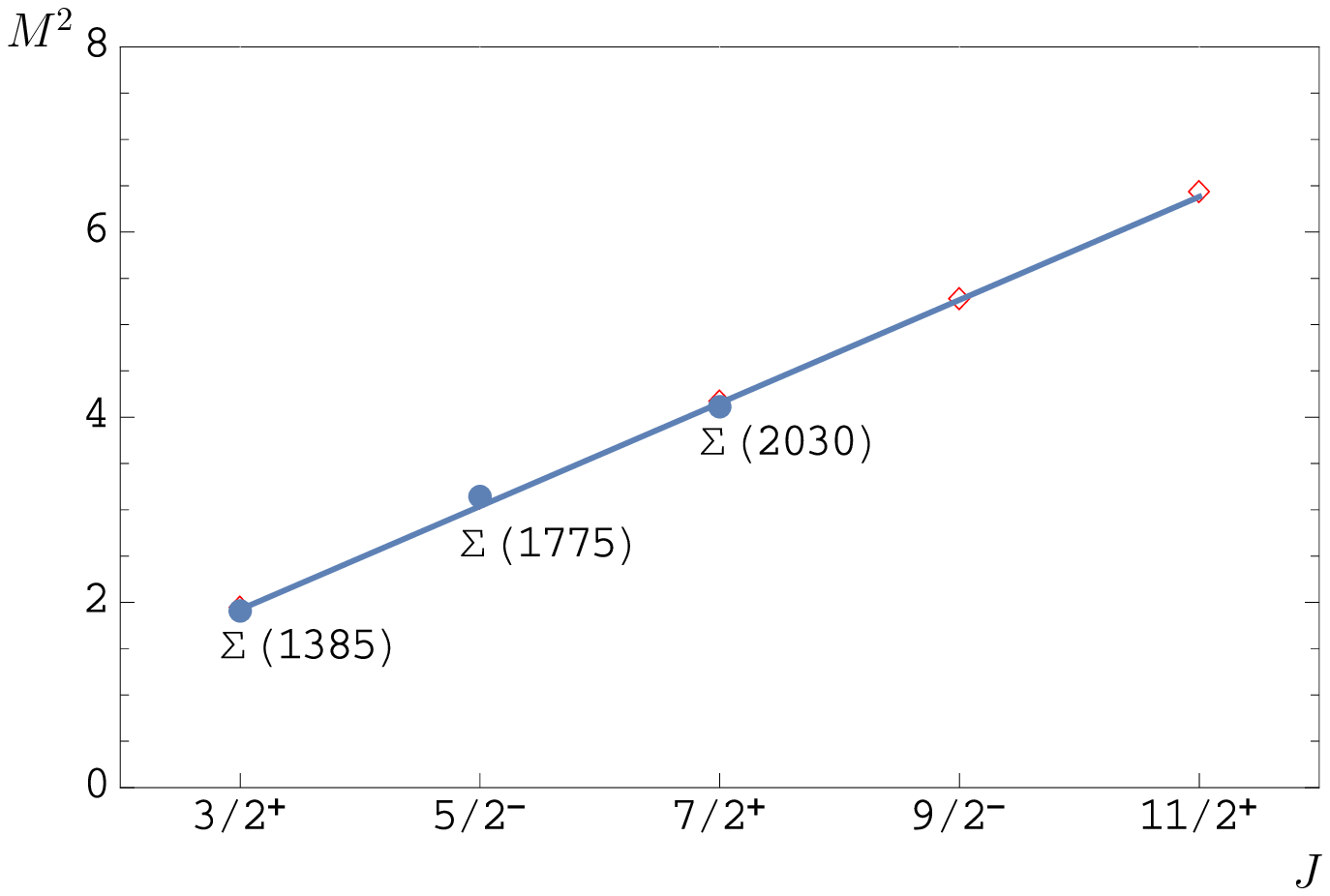}\\
\vspace*{-0.3cm}\hspace*{0.5cm} (a) \hspace*{7.5cm} (b) 

\caption{\label{fig:sigma_j} Same as in Fig.~\ref{fig:lambda_j} for
  the $\Sigma$ baryons. }\vspace*{0.5cm}
\end{figure}

\begin{figure}
 \includegraphics[width=8cm]{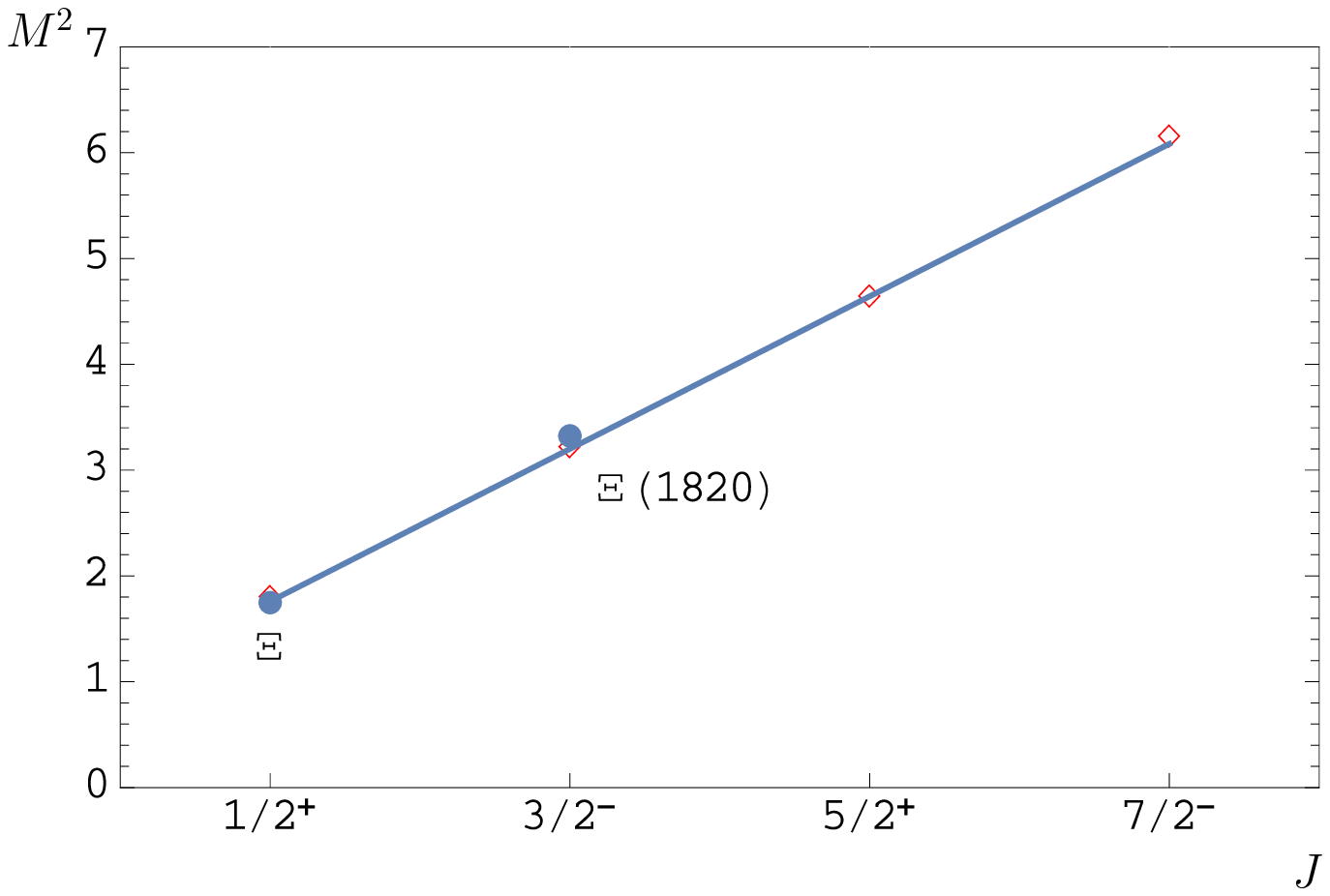}\ \ \ \includegraphics[width=8cm]{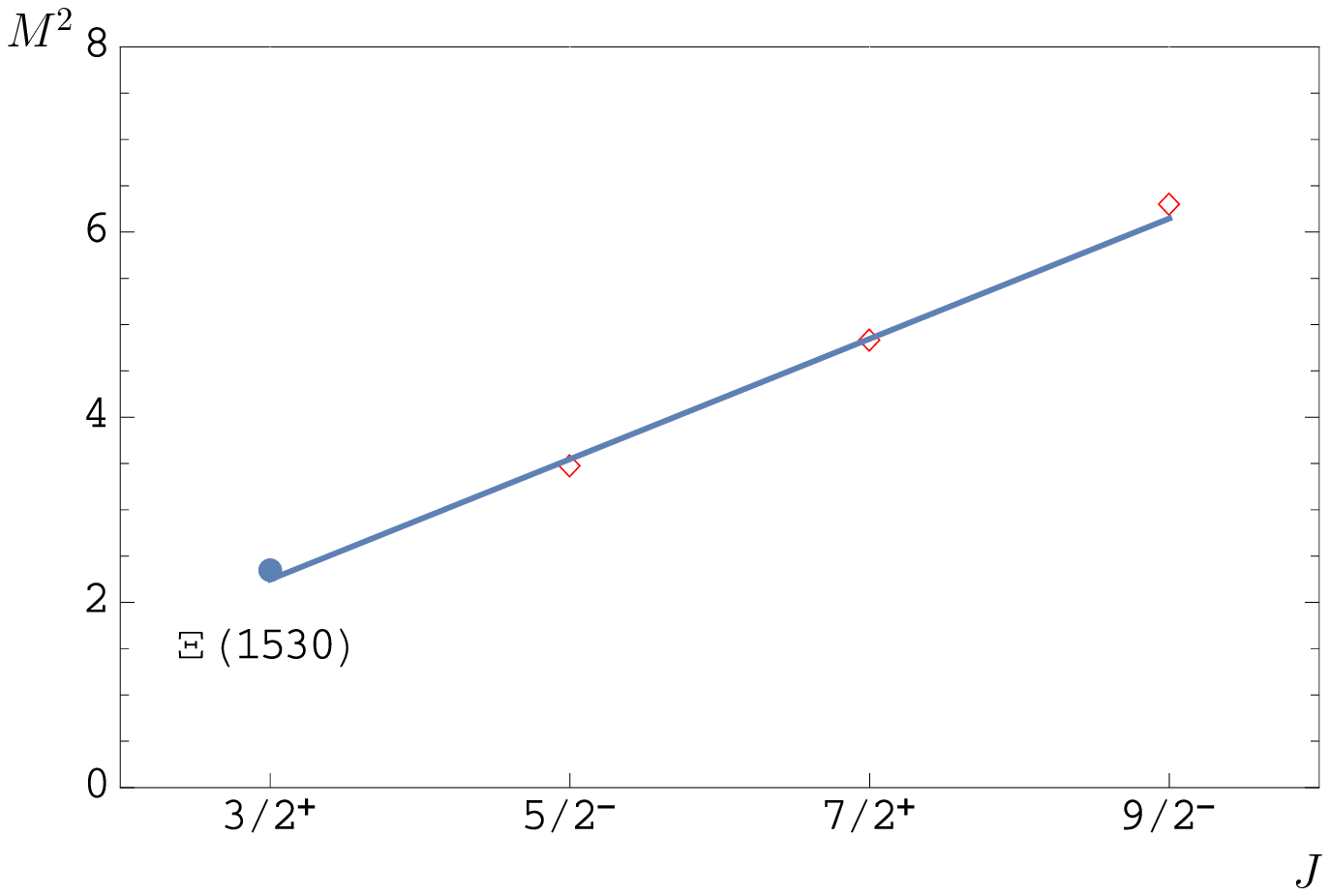}\\
\vspace*{-0.3cm}\hspace*{0.5cm} (a) \hspace*{7.5cm} (b) 

\caption{\label{fig:xi_j} Same as in Fig.~\ref{fig:lambda_j} for
  the $\Xi$ baryons. } 
\end{figure}

In Figs.~\ref{fig:lambda_j}-\ref{fig:xi_j} we plot the Regge trajectories in
the ($J, M^2$) plane for strange baryons with natural
($P=(-1)^{J-1/2}$) and unnatural ($P=(-1)^{J+1/2}$) parities. The masses calculated in our
model are shown by diamonds. Available experimental data are given by
dots with error bars and corresponding baryon names. 
Straight lines were obtained by the
$\chi^2$ fit of calculated values. The fitted slopes
and intercepts of the Regge trajectories are given in
Table~\ref{tab:rtj}. We see that the calculated
strange baryon masses lie on the linear trajectories.

\begin{table}
  \caption{Fitted parameters $\alpha$, $\alpha_0$ for the slope and intercept of the $(J,M^2)$ Regge
    trajectories of strange baryons.} 
  \label{tab:rtj}
\begin{ruledtabular}
\begin{tabular}{ccc@{\ \ \ }||ccc}
Baryon&
$\alpha$ (GeV$^{-2}$)& $\alpha_0$&Baryon&$\alpha$
(GeV$^{-2}$)&$\alpha_0$\\
\hline
$\Lambda$ $(\frac12^+)$&$0.923\pm0.016$&$-0.648\pm0.057$&
$\Lambda$ $(\frac12^-)$&$0.732\pm0.018$&$-0.951\pm0.074$\\ 
$\Sigma$ $(\frac12^+)$&$0.799\pm0.029$&$-0.676\pm0.100$&
$\Sigma$ $(\frac32^+)$&$0.897\pm0.010$&$-0.225\pm0.037$\\ 
$\Xi$ $(\frac12^+)$&$0.694\pm0.007$&$-0.721\pm0.024$&
$\Xi$ $(\frac32^+)$&$0.769\pm0.032$&$-0.249\pm0.098$\\ 
&&&$\Omega$ $(\frac32^+)$&$0.712\pm0.002$&$-0.504\pm0.007$\\
\end{tabular}
 \end{ruledtabular}
\end{table}

The natural parity $\Lambda$ Regge trajectory is the best studied
experimentally. There are five well established (four 4-star and one
3-star) states \cite{pdg} on this trajectory. The masses of these states
calculated in our model agree well with data. Using the constructed
Regge trajectory we can predict the mass of the $\frac{11}2^-$
$\Lambda$ state to be about 2605~MeV (see Table~\ref{lambdamin}). This
state could  contribute to the $\Lambda(2585)$ bumps observed
with the mass $\approx 2585$~MeV \cite{pdg}.  
Each of the $\Sigma$ Regge trajectories contains three well
established states \cite{pdg}, well fitting to the strait lines.   
Other trajectories are less motivated experimentally 
and contain at most two well
established states.

Using the values of the slopes and intercepts of the Regge
trajectories of the $\frac32^+$ strange baryons we can test the
validity of the relations between them proposed in the literature (see
e.g. \cite{k,bg,gww} and references therein). It is easy to check that
the additivity of inverse slopes
 \begin{equation}
  \label{eq:aisl}
  \frac1{\alpha(\Sigma^*)}+\frac1{\alpha(\Omega)}=\frac2{\alpha(\Xi^*)},
\end{equation}
factorization of slopes
\begin{equation}
  \label{eq:fsl}
  \alpha(\Sigma^*)\alpha(\Omega)=\alpha^2(\Xi^*),
\end{equation} 
and additivity of intercepts
\begin{equation}
  \label{eq:aint}
  \alpha_0(\Sigma^*)+\alpha_0(\Omega)=2\alpha_0(\Xi^*),
\end{equation}
are well satisfied. Indeed, in the left hand side of
Eq.~(\ref{eq:aisl}) we get $2.52\pm0.02$ and in the right hand side
$2.60\pm0.11$; for Eq.~(\ref{eq:fsl}) the corresponding values are
$0.639\pm0.010$ and $0.592\pm0.050$, while for Eq.~(\ref{eq:aint}) they
are $-0.729\pm0.044$ and $-0.498\pm0.196$.

We can also compare the calculated slopes of the strange baryon Regge
trajectories with our previous results for the slopes of  heavy baryons
\cite{hbarRegge} and light mesons \cite{lregge}. Such comparison
shows that the strange baryon slopes lie just in between the corresponding
slopes of light mesons and charmed baryons. Moreover they follow the
same pattern as the slopes of heavy baryons: the slope decreases with the
increase of the diquark mass as well as with the increase of the parent
baryon mass.  

\section{Conclusions}
\label{sec:concl}

The mass spectra of strange baryons were calculated in the framework
of the relativistic quark model based on the quasipotential
approach. The quark-diquark picture, which had been previously
successfully 
applied for the investigation of the spectroscopy of heavy baryons
\cite{hbar,hbarRegge}, was extended to the strange baryons. Such
approach allows one to reduce very complicated relativistic three-body
problem to the subsequent solutions of two two-body problems. It is
assumed that the baryon is the bound quark-diquark system, where two
quarks with equal constituent masses form a diquark. The diquarks are
not treated to be the point-like objects. Instead their internal
structure is taken into account by the introduction of the form
factors  expressed in terms of the diquark wave functions.  The diquark
masses and form factors were calculated using the solutions of the
relativistic quasipotential equation with the kernel which
nonperturbatively accounts for the relativistic effects. It was found
that for the correct description of the strange baryon mass spectra it is
necessary to consider not only the ground state scalar and 
axial vector diquarks, as we did in our previous study of
heavy baryon spectroscopy \cite{hbarRegge}, but also their first
orbital and radial excitations. The ground state and excited baryon
masses were obtained by solving  the relativistic quark-diquark
quasipotential equation. Note that in our analysis we did
not make any new assumptions about the quark interaction in baryons or
introduce any new parameters. The values of all parameters were taken
from previous considerations of meson properties. This
significantly increases the reliability and predictive power of our
approach.  The masses of strange baryons were calculated
up to rather high 
orbital and radial excitations. This allowed us to construct the Regge
trajectories which were found to be linear. The validity of the
proposed relations between the Regge slopes and intercepts was tested.

The obtained results were compared with available experimental data \cite{pdg}
and previous predictions within different theoretical approaches \cite{ci,lmp,mps,sf,elms}. We 
found that all 4- and 3-star states of strange baryons with established
quantum numbers are well reproduced in our model as well as most of the
2- and 1-star states. Possible candidates for the experimentally
observed states with unknown quantum numbers can be identified. We
emphasize that the experimental mass of the
$\Lambda(1405)$ is naturally reproduced in our model, while its rather
low mass presents some difficulties for most of the three-quark models
\cite{ci,lmp,mps}.  It is
necessary to note that our quark-diquark picture predicts less excited
states of strange baryons than the three-body approaches. The
distinctions become apparent for higher baryon
excitations. However the number of predicted strange baryon states
still significantly exceeds the number of observed ones. Thus the
experimental determination of the quantum numbers of the already observed $\Xi$ and $\Omega$
excited states as well as the further search for the missing excited
states of strange baryons
represents highly promising and important problem.    

\acknowledgments
The authors are grateful to 
D.~Ebert,  V. A. Matveev  and V. I. Savrin  
for  useful discussions.

\end{document}